\documentclass[%
 reprint,
superscriptaddress,
 amsmath,amssymb,
 aps,
pra,
]{revtex4-2}
\usepackage{lipsum}
\usepackage{graphicx}
\usepackage{dcolumn}
\usepackage{bm}
\usepackage{hyperref}
\usepackage{xcolor}
\usepackage[dvipsnames]{xcolor}

\begin{document}

\title{Observation of Photon Bunching in Broadband Multimode Thermal Light Enabled by Temporal Imaging}
\author{Filip So\'snicki}
\email{filip.sosnicki@fuw.edu.pl}
\affiliation{Faculty of Physics, University of Warsaw, Pasteura 5, 02-093 Warszawa, Poland}
\affiliation{Paderborn University, Integrated Quantum Optics, Institute for Photonic Quantum Systems (PhoQS), Warburger Str.~100, 33098 Paderborn, Germany}

\author{Micha\l{}  Miko\l{}ajczyk}
\author{Sanjay Kapoor}
\affiliation{Faculty of Physics, University of Warsaw, Pasteura 5, 02-093 Warszawa, Poland}

\author{Dmitri B. Horoshko}
\affiliation{Univ. Lille, CNRS, UMR 8523 -- PhLAM -- Physique des Lasers Atomes et Mol\'ecules, 59000 Lille, France}
\affiliation{Institut f\"ur Quantenoptik, Universit\"at Ulm, Ulm D-89081, Germany}

\author{Mikhail I. Kolobov}
\affiliation{Univ. Lille, CNRS, UMR 8523 -- PhLAM -- Physique des Lasers Atomes et Mol\'ecules, 59000 Lille, France}

\author{Micha\l{}  Karpi\'nski}
\affiliation{Faculty of Physics, University of Warsaw, Pasteura 5, 02-093 Warszawa, Poland}

\date{\today}

\begin{abstract}
Photons emitted by incoherent sources exhibit temporal bunching, a fundamental manifestation of Bose-Einstein statistics first observed by Hanbury Brown and Twiss (HBT). While intensity correlation measurements originated in astronomy and are now central across quantum technologies, biophysics, and spectroscopy, observing temporal bunching in broadband thermal light remains limited by detector temporal resolution. Conventional approaches -- such as streak cameras or optical upconversion -- introduce severe trade-offs in efficiency, bandwidth, or setup versatility. Here, we propose and demonstrate a third route: pre-detection temporal magnification of the optical field prior to measuring its intensity autocorrelation function in a conventional HBT setup. Using an electro-optic time lens with a temporal magnification of $660$, we resolve the temporal intensity autocorrelation of broadband thermal light generated via non-degenerate spontaneous parametric down-conversion with a bandwidth of $250$~GHz. This enables direct observation of thermal photon bunching on sub-$3.5$ picosecond timescales without narrow spectral filtering, establishing pre-detection temporal imaging as a general platform for broadband statistical optics, with applications spanning ultrafast quantum optics, biophysical imaging, and potentially stellar intensity interferometry.
\end{abstract}

\maketitle

\section{Main}

The statistical properties of light play an essential role in both classical and quantum optics, as they have fundamental implications and diverse applications \cite{Mandel, Stiller14,  Shirai17, Laiho22,  Tan23}. 
One of the key features of the thermal light statistics is the effect of photon bunching, meaning that two photons are more likely to be detected close together in space and time than farther apart \cite{Mandel}. Observation of spatial photon bunching in starlight by Hanbury Brown and Twiss (HBT) \cite{HanburyBrown56a, HanburyBrown56} led to the emergence of intensity stellar interferometry, allowing one to measure the angular size of distant stars far beyond the capabilities of conventional astrometry \cite{Karl22}. Spatial, temporal, and spectral bunching of photons in thermal and pseudothermal light is widely used in ghost imaging \cite{Bennink02,Gatti04,Liu14}, temporal ghost imaging \cite{Ryczkowski16}, and ghost spectroscopy \cite{Janassek18}. The opposite effect of photon antibunching has no classical explanation. The observation of antibunching of photons emitted by a single atom \cite {Kimble77} was the first direct proof of the quantum nature of light. Today, photon antibunching measurement is a routine procedure for the characterization of single-photon sources such as quantum dots \cite{Arakawa20}, single molecules \cite{Toninelli21}, and diamond colour centres \cite{Doherty13}, which are widely developed for applications in photonic quantum technologies. Antibunching is also used in microscopy to improve resolution beyond the diffraction limit \cite{Tenne19}. 

The temporal statistics of photons in optical fields are typically characterized by the normalized second-order autocorrelation function \cite{Glauber63g2}
\begin{equation}\label{g2}
    g^{(2)}(\tau) = \frac{\langle :\hat I(t+\tau)\hat I(t):\rangle}{\langle\hat I(t+\tau)\rangle \langle\hat I(t)\rangle},
\end{equation}
where $\hat I(t)=\hat E^{(-)}(t)\hat E^{(+)}(t)$ is the field intensity operator, $\tau$ is the delay time, $\hat E^{(+)}(t)$ and $\hat E^{(-)}(t)$ are the positive and negative frequency parts of the field operator, respectively, and the colon stands for time and normal ordering. 
The value of this function at zero delay, $g^{(2)}(0)$, is of great importance in quantum optics as it captures the statistical properties of light, revealing the physical mechanism of its generation. It is $1$ for coherent laser light, whereas any classically fluctuating field has $g^{(2)}(0)>1$, which is a signature of photon bunching. For the thermal light obeying the Bose-Einstein statistics, the value of $g^{(2)}(0)$ is exactly 2, providing thus a powerful tool for distinguishing the thermal sources of light. Values below unity correspond to the non-classical effect of anti-bunching and allow one to identify single quantum emitters. 

\begin{figure}
    \includegraphics[width=0.9\linewidth]{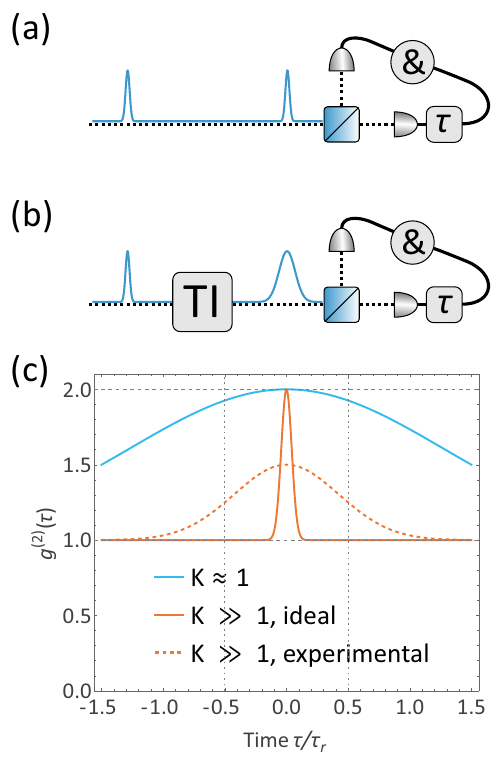}
    \caption{ {\bf Time-resolved second-order autocorrelation measurement.} (a) Hanbury Brown and Twiss (HBT) interferometer comprising a 50:50 beamsplitter, two detectors, and a coincidence counter. (b) Temporal imaging (TI) preceding the HBT interferometer magnifies the signal timescale and, consequently, the time-resolved second-order autocorrelation, which reduces the required detector temporal resolution. (c) Normalized second-order coherence $g^{(2)}(\tau)$ of thermal light for different numbers of modes $K$ of the measured signal. For $K\approx1$, the broad correlation feature is resolved (solid blue line), while increasing $K$ progressively narrows the feature (solid orange line). When its width falls below the detector resolution $\tau_{r}$, temporal averaging reduces $g^{(2)}(0)$ below 2 (dashed orange line). Temporal imaging allows features shorter than the detector resolution to be resolved by stretching their temporal profile. The grey vertical dotted lines mark the detector resolution $\tau_r$, which is the shortest experimentally measurable duration.
    }
    \label{fig:idea}
\end{figure}

Experimentally, the value of $g^{(2)}(0)$  should be measured in time intervals much shorter than the coherence time of light $\tau_c$, since at times much longer than $\tau_c$ the intensity correlation vanishes and $g^{(2)}(\tau)$ tends to $1$. For continuous-wave light, $\tau_c$ is inversely proportional to the spectral bandwidth, which imposes a strict limit on the measured signal. For pulsed light, coherence modes can be found via the Mercer expansion \cite{Mandel} and $\tau_c$ also diminishes with the growing number of these modes $K$. While the temporal resolution of modern commercially available photodetectors is typically limited to $>10$~ps (though resolutions down to $\sim 3$~ps have been reported with superconducting nanowire single-photon detectors \cite{Korzh2020}), direct measurement of $g^{(2)}(\tau)$ around the origin is effectively restricted to fields with a bandwidth below $100$~GHz. For broader bandwidths, the detector resolution becomes insufficient to resolve the relevant temporal features, resulting in time-averaging effects, as illustrated in Fig.~\ref{fig:idea}~(c). This resolution limit prevents measuring autocorrelation at fundamentally interesting, very short timescales and hinders the practical application of correlation measurements through the narrow spectral bandwidth requirement. 

In this study, we break this fundamental limitation by utilizing temporal imaging \cite{torres11,salem13}—a coherent manipulation of light that allows us to modify the measurement timescale by over two orders of magnitude. This enables us to measure the maximum bunching for spectrally broadband multimode thermal signals without resorting to spectral filtering or narrowband light generation. We experimentally demonstrate $g^{(2)}(0) = 2$ for thermal signals with a spectral bandwidth two orders of magnitude larger than previous direct demonstrations \cite{Tan14, Mika18}. Our results will enable improved understanding of time-dependent photon statistics, especially by opening the way towards autocorrelation measurements at optical single-cycle time scales,  as well as new applications of bunched thermal light \cite{TorresCompany10,BeneaChelmus16, Goetz17, Raino18}.

Prior to this work, overcoming the detector resolution limit was largely dependent on source manipulation or alternative detection schemes. A well-known method to bypass this limitation is to emulate temporally single-mode and spatially multimode thermal light using rotating ground glass plates to scatter coherent radiation \cite{arecchi65}, which produces ``pseudothermal'' fluctuations on easily measurable, slow timescales. For true thermal sources, such as the Sun, an incandescent lamp, or single-pass spontaneous parametric down-conversion (SPDC), the measurement has only been possible with ultra-narrowband spectral filtering \cite{Tan14, Deng19, ruppel2023}. Exceptions include thermal sources that are inherently narrowband enough to allow for a full measurement of the function $g^{(2)}(\tau)$, such as hot atomic vapours \cite{Mika18}, some stellar objects \cite{Tan17}, and ultra-narrowband SPDC in resonant structures \cite{Luo2017}. Finally, while macroscopic chaotic emission from semiconductor lasers can exhibit broadband fluctuations (tens of GHz) where temporal bunching is fully resolved \cite{wang18}, the underlying gain dynamics fundamentally differs from the true Bose-Einstein statistics of spontaneous emission. Nonetheless, relying on ultra-narrowband light means that the majority of true broadband thermal sources cannot presently be faithfully characterized by their true photon statistics using these methods.

\begin{figure*}
    \includegraphics[width=\textwidth]{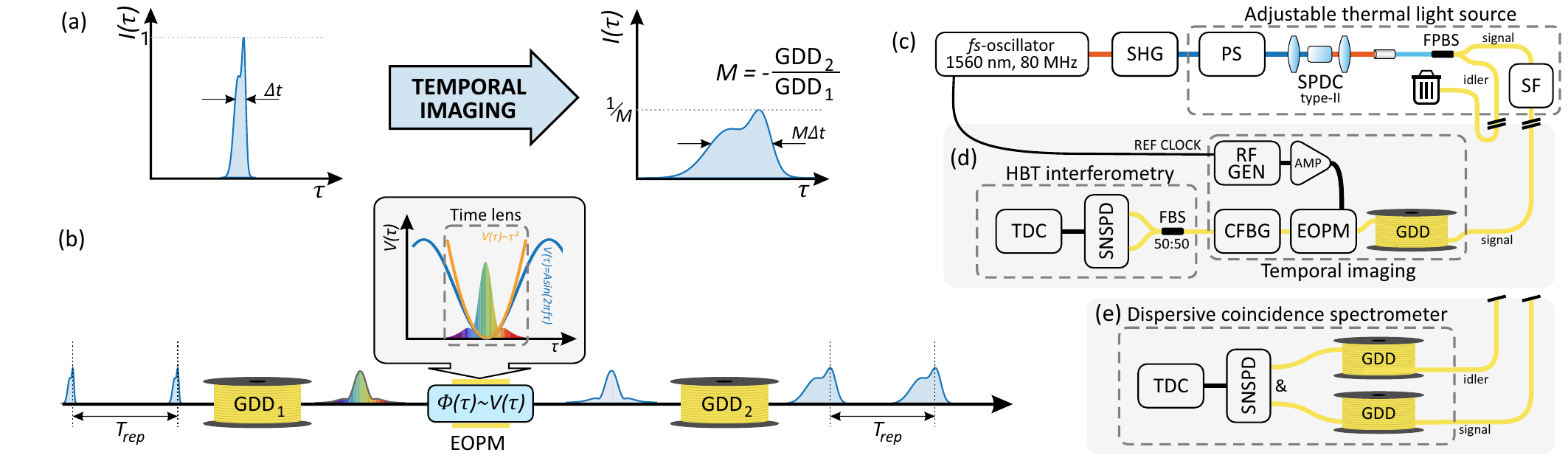}
    \caption{ {\bf Temporal imaging and experimental setup.}
        (a) Temporal imaging reshapes an optical pulse by increasing its duration while preserving its temporal profile. 
        (b) The temporal imaging system uses two group-delay dispersions (GDDs) together with temporal phase modulation. Each pulse is stretched by the first GDD, after which a quadratic temporal phase is applied. The second GDD then further stretches the pulse, retrieving the pulse's magnified temporal envelope. It is a pulse-by-pulse process that temporally magnifies the pulses' envelope without changing the pulse repetition period. 
        (c, d, e) Schematic of the experimental setup.
        (c) An amplified $f\!s$-oscillator produces laser pulses that are frequency-doubled via second-harmonic generation (SHG) and subsequently fed into an adjustable thermal light source based on spontaneous parametric down-conversion (SPDC). The SPDC cross-polarized and spectrally non-degenerate signal-idler photon pairs are separated using a fibre polarization beam-splitter (FPBS). The idler photons are discarded, leaving a thermal state in the signal field. The thermal state mode number is controlled by the pump's pulse shaper (PS).
        (d) To measure the time-imaged second-order correlation function, we apply a tunable spectral filter (SF) to the signal photons to improve the suppression of the idler photons. The signal photons undergo temporal imaging, consisting of temporal stretching in a fibre-based GDD medium, application of a temporal phase, and subsequent temporal stretching in a chirped fibre Bragg grating (CFBG). The quadratic temporal phase is approximated by an amplified (AMP) single-tone voltage waveform generated by a radio-frequency generator (RF GEN) locked to the laser repetition rate. Finally, the temporally magnified signal photons are measured using a fibre-based Hanbury Brown–Twiss (HBT) interferometer consisting of a fibre 50:50 beamsplitter (FBS), followed by superconducting nanowire single-photon detectors (SNSPDs) and a time-to-digital converter (TDC).
        (e) To characterize the number of modes in the SPDC output fields, we separate the signal and idler photons using the FPBS and use the same settings of the signal spectral filter (SF) as above. We then apply a large GDD separately to the signal and idler photons and measure their coincidences to obtain the joint spectral intensity (JSI), from which we determine the number of modes (Methods).
    }
    \label{fig:setupTI}
\end{figure*}

Several routes were pursued to bypass this difficulty. One route consists of using streak cameras \cite{Streak} that have picosecond temporal resolution but an extremely low repetition rate, which makes them not very attractive for the characterization of relatively weak fields. Another route consists of using nonlinear optical processes, such as two-photon absorption in semiconductors \cite{Boitier09} or doubly-gated upconversion with variable gate separation \cite{Delteil19}. Both approaches of this route lack the versatility of the HBT setup: the two-photon absorption is a resonant effect available only for a fixed set of wavelengths, and the gated upconversion has a very low quantum efficiency (of the order $10^{-5})$ and, similar to the streak cameras, is not very attractive for the characterization of relatively weak fields.  

\begin{figure*}
    \includegraphics[width=\textwidth]{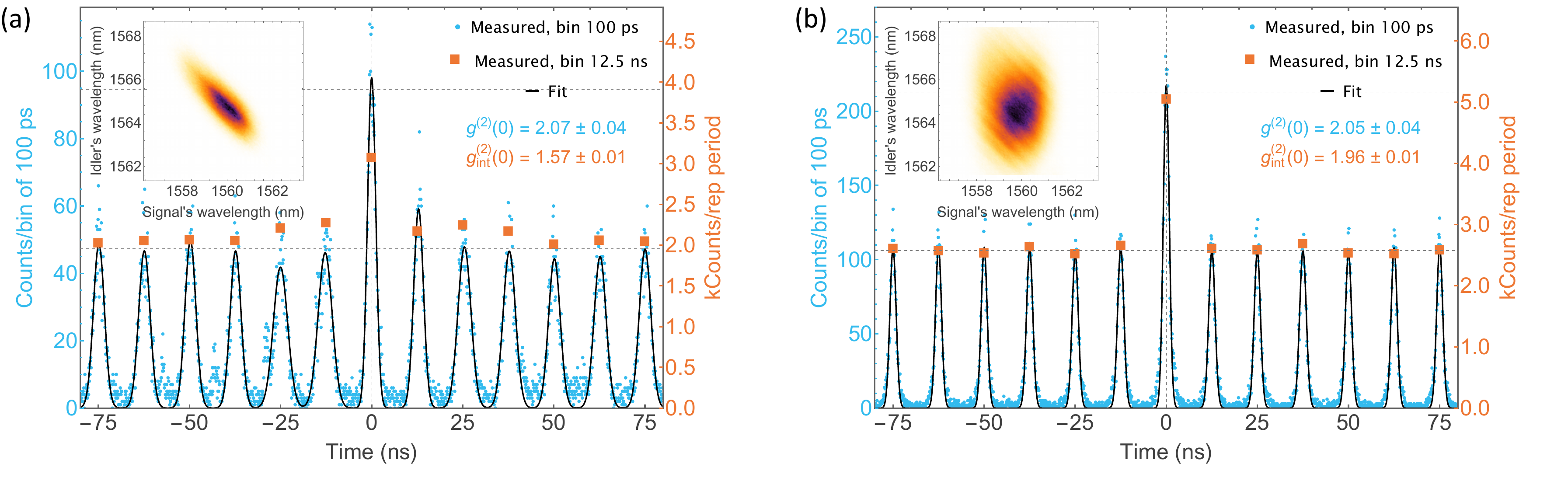}
    \caption{{
        \bf Time-magnified photon count histograms.}
        660-fold temporal-magnified time-delayed coincident photon count histograms when pumping the SPDC with a spectral bandwidth of (a) $0.2$~nm and (b) $0.9$~nm, corresponding to Schmidt numbers $K = 1.68$ and $K = 1.02$, respectively. The temporally magnified measurement points were binned into $100$~ps bins (corresponding to $150$~fs through the 660 temporal magnification factor; blue) for high temporal resolution and into $12.5$~ns bins (corresponding to the pulse repetition period, which is not temporally magnified; orange) for low temporal resolution. Notice the central peak is narrower as compared to the others in (a), while the same in (b), which shows the influence of shorter coherence time due to the multimodal structure of light. The insets show the joint spectral intensities of SPDC-generated photon pairs used for the thermal state generation. 
    }
    \label{fig:g2graphs}
\end{figure*}

In addition to the limitations mentioned, none of these approaches, to the best of our knowledge, has enabled experimental observation of the time-resolved value $g^{(2)}(0) = 2$ for spectrally broadband thermal optical fields predicted by the theory \cite{Mandel,Glauber63g2}. As a consequence, the majority of thermal sources cannot be presently identified by their photon statistics. 

Our technique enhances the conventional HBT setup \cite{HanburyBrown56a} by applying the aforementioned temporal imaging to the optical field prior to measurement. Temporal imaging of optical fields is a well-established technique that realizes temporal magnification of the optical signal, similar to spatial magnification by a conventional lens. It is achieved by combining the propagation of an optical pulse in a second-order dispersive medium, a parabolic time-dependent phase modulation known as the time lens \cite{kolner89}, and subsequent propagation in another second-order dispersive medium. By appropriately choosing the values of the group delay dispersion (GDD) of the media and the phase modulation amplitude, one can achieve temporal magnification of the input signal, with the magnification factor given by the ratio of the GDDs of the two dispersive media. Although developed mainly for strong classical fields, temporal imaging was also successfully applied to non-classical and single-photon fields \cite{Lavoie13,Karpinski17}, and the corresponding quantum theory of this technique has been developed \cite{Patera18,Shi20}. The effect of temporal imaging is presented schematically in Figure \ref{fig:setupTI}a.

In our experimental implementation, the pulsed thermal light -- one of the two twin beams generated in non-degenerate SPDC -- is produced at a fixed repetition rate of $80$~MHz. The temporal magnification is achieved with the same repetition rate. This results in a temporal magnification within each repetition period of $12.5$ ns individually, as shown schematically in Fig. \ref{fig:setupTI}b. We implement the time lens using electro-optical phase modulation and achieve a temporal magnification of $660$ by the appropriate choice of the two dispersive media (Methods). We expand an approximately $3$-ps-long (full-width at half maximum, FWHM) single-photon pulse at the centre of each repetition period to a $660$ times longer duration of about $2$~ns (FWHM). The resolution of our temporal imaging setup is better than $3.4$~ps (see Methods), allowing us to measure $g^{(2)}(\tau)$ with this resolution. The theoretical treatment of this technique can be found in Ref.\ \cite{horoshko25}.

The $g^{(2)}(\tau)$ function is measured using the standard HBT approach, by directing the time-magnified optical signal to a $50/50$ beam splitter, followed by two time-resolved photon-counting detectors and registering the number of coincident counts with respect to the time interval $\tau$ between clicks at two detectors, as shown in Figs. \ref{fig:setupTI}c and \ref{fig:setupTI}d. The histogram of coincident counts with respect to $\tau$, after appropriate normalization, yields a measurement of $g^{(2)}(\tau)$ (see Methods). In most previous works dealing with SPDC, where the temporal resolution was less than the coherence time of the field, only the time-integrated function $g_\text{int}^{(2)}$ was measured, which is less than 2 for a multimode thermal light \cite{christ11}. In contrast, our time-resolved measurement shows that the value of $g^{(2)}(0)$ is close to 2 within the measurement error independently of the number of thermal light modes. Our technique paves the way for precise statistical characterization of broadband light sources and for new applications of bunched thermal light.

\section{Time-resolved autocorrelation of broadband SPDC}

To comprehensively demonstrate the measurement of temporal fluctuations on single-picosecond time scales, one requires a source of thermal light where the number of spectral modes -- and therefore the coherence time -- can be precisely tuned independently of the overall optical bandwidth. SPDC serves as an ideal model system for this purpose.

Its primary advantage over conventional thermal sources is the easily controllable mode number. In SPDC, a pulsed pump beam illuminates a second-order nonlinear crystal, generating two distinct, correlated optical fields, typically labelled as signal and idler \cite{Mandel}. When observed in isolation -- effectively tracing out the unmeasured field --  each field exhibits thermal photon statistics, where the number of modes is determined by the spectral-temporal correlations between the signal and idler \cite{horoshko25, christ11, Faruque19}. 
We experimentally tune these correlations by controlling the spectral bandwidth of the pump pulses, using a home-built femtosecond pulse shaper (PS, Fig.~\ref{fig:setupTI}c and Methods). The resulting spectro-temporal correlations, captured by the joint spectral intensity (JSI) of the signal and idler, are experimentally monitored using the dispersive Fourier transform technique in the coincident single-photon counting regime (Fig. \ref{fig:setupTI}e, Methods and Supplementary Section S1).

Mathematically, these spectro-temporal correlations are captured by the Schmidt number $K$ \cite{Law00}, which corresponds to the number of correlated modes and, consequently, to the number of coherence modes in the individually observed thermal signal field. Although $K$ is formally defined via the complex joint spectral amplitude (JSA) \cite{horoshko25}, we determine its value directly from our measured JSIs by assuming a real-valued JSA (i.e., flat joint spectral phase) -- an assumption well-justified for our collinear SPDC source (Methods). While SPDC fundamentally permits fully decoupling the spectral bandwidth and the number of modes during source design, in our experiment, we vary the number of modes directly via the SPDC pump bandwidth. By tuning the pump spectral bandwidth between $0.20$~nm and $1.00$~nm (FWHM), we continuously vary the Schmidt number from $1.68$ (multimode thermal light) to $1.02$ (near-pure single-mode thermal light), reaching a signal spectral bandwidth of $2.00$~nm ($246.5$~GHz) in the single-mode limit (see Supplementary Section S1B for further details). 

In classical statistical optics, time-averaging over $K$ independent temporal modes suppresses the observed intensity correlation contrast to $g^{(2)}(0) = 1 + 1/K$ \cite{Loudon-book}. For our terahertz-order-bandwidth multimode signal, the single-picosecond coherence time is shorter than the response time of single-photon detectors, causing significant mode-averaging that blurs the observed photon bunching. By stretching the optical field in time prior to detection, our temporal magnifier ensures that the detector's timing jitter samples well within a single magnified coherence interval. This removes the time-averaging bottleneck, allowing us to directly resolve $g^{(2)}(0) \rightarrow 2$ regardless of $K$. Also, we couple the signal and idler fields to single-spatial-mode optical fibres, to remove any influence of spatial modes or correlations on $K$. 

To measure $g^{(2)}(\tau)$ for the isolated signal beam (leaving the idler field undetected), we implemented the experimental configuration shown in Fig.~\ref{fig:setupTI}(d), incorporating a $660\times$ temporal magnification stage \cite{kolner89}
preceding a Hanbury Brown and Twiss (HBT) interferometer (Methods). The HBT setup comprised a balanced fibre beam splitter feeding two superconducting nanowire single-photon detectors (SNSPDs; $\sim 30\text{ ps}$ temporal resolution), with photodetection events from both channels registered by a high-resolution time-tagger (Methods). 

Fig.\ \ref{fig:g2graphs} displays the measured coincidence count histograms as a function of delay $\tau$ between the photodetection events. Figure \ref{fig:g2graphs}(a) shows the multimode case of $K=1.68$, while Fig.\ \ref{fig:g2graphs}(b) presents the approximately single-mode case of $K=1.02$. Note that the temporal magnification is applied strictly within each individual pulse repetition period, leaving the inter-pulse spacing unscaled. Under high-resolution temporal imaging ($100~\mathrm{ps}$ post-magnification binning, blue points), we observe normalized zero-delay second-order autocorrelation values, $g^{(2)}(0)$, of $2.08 \pm 0.04$ and $2.04 \pm 0.04$ respectively -- both agreeing with the theoretical value of $2.00$ within up to two standard deviations. The non-zero delay peaks correspond to uncorrelated light from adjacent laser pulses and serve as a natural reference for experimental $g^{(2)}(\tau)$ normalization.

These results represent the first direct measurement of full thermal bunching, ($g^{(2)}(0) = 2$), in a broadband thermal field,  obtained by increasing the temporal resolution. This provides direct experimental validation of the foundational statistical optics principle that chaotic fields exhibit maximum $g^{(2)}(0) = 2$ contrast when resolved on time scales shorter than their intrinsic coherence intervals \cite{Loudon-book}. 

\begin{figure}
    \includegraphics[width=\linewidth]{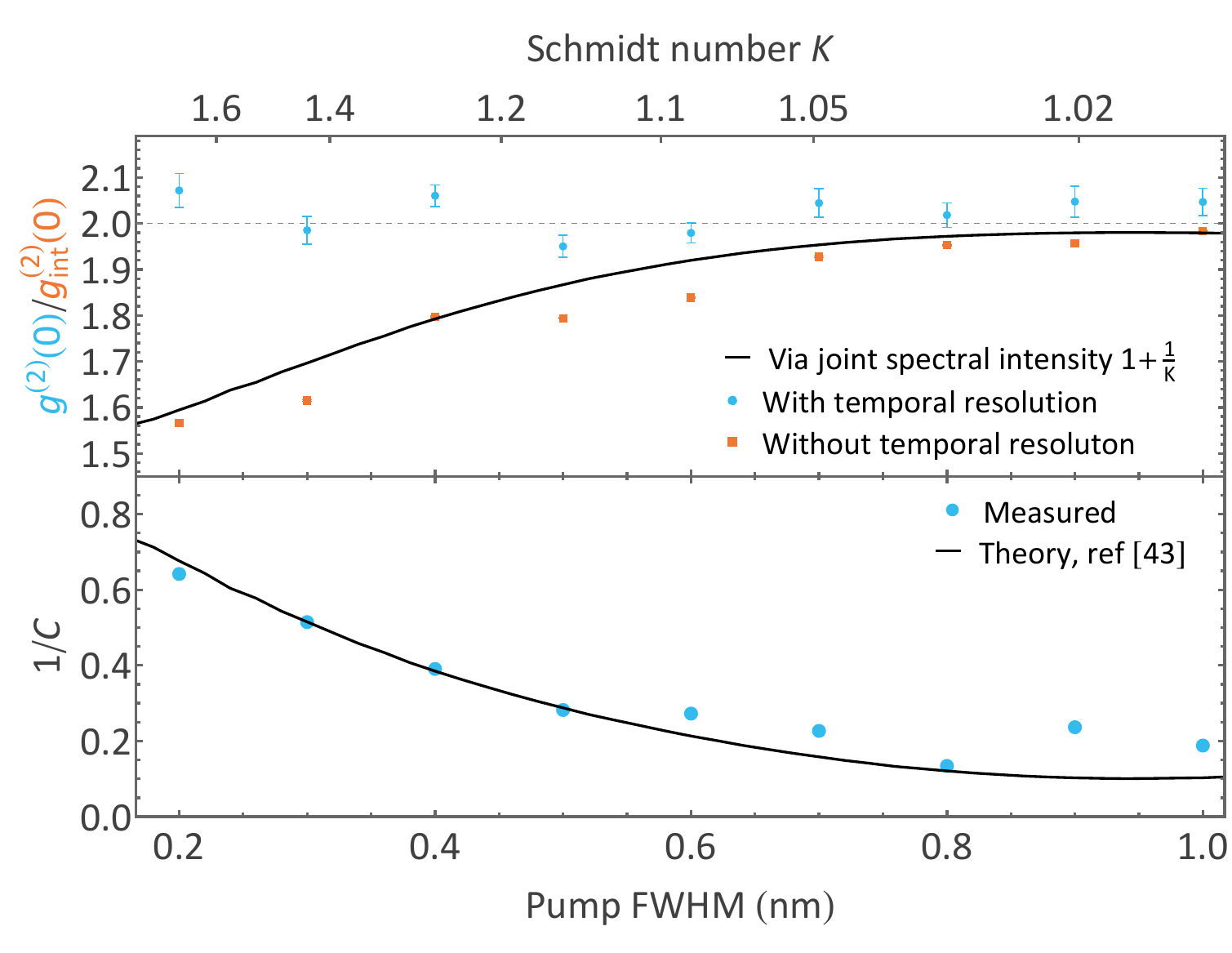}
    \caption{ 
        {\bf Scaling with the number of modes.} Top: Zero-delay values of the second-order coherence function $g^{(2)}(\tau = 0)$ for a range of the effective number of modes $K$, controlled by varying the pump spectral bandwidth, measured with high (blue) and low (orange) temporal resolution. The dashed horizontal line indicates the maximal value of 2 expected for the thermal state, while the black solid line shows the detector temporal averaging effect as $1+\frac{1}{K}$.
        Bottom: Ratio $\Delta t_M/\sqrt{2}\Delta \tau_M$ after temporal imaging, corresponding to the inverse global degree of coherence $1/C$, for varying effective numbers of modes $K$. The solid black line is the theoretical prediction derived in ref.~\cite{horoshko25}. 
    }
    \label{fig:results}
\end{figure}

To highlight the impact of high temporal resolution enabled by temporal imaging, we also calculate the so-called time-integrated (or pulsed) normalized second-order autocorrelation $g^{(2)}_\mathrm{int}$,  which is often used as a measurement tool in quantum optics \cite{christ11}, by integrating photon counts across entire pulse periods ($12.5\text{ ns}$ binning; orange points in Fig.\ \ref{fig:g2graphs} and the top panel of Fig.\ \ref{fig:results}). Figure \ref{fig:results}a compares these time-integrated values against our time-resolved measurements across the full range of $K$. Without temporal magnification, $g^{(2)}_{\text{int}}$ follows the well-established mode-averaging curve $1 + 1/K$, dropping significantly as $K$ increases. In contrast, our temporally magnified measurements consistently yield $g^{(2)}(0) \approx 2$ across all mode numbers.

Finally, temporal magnification enables us to track how the duration of the central autocorrelation peak scales relative to the pump bandwidth. In the bottom panel of Fig.\ \ref{fig:results}, we plot the inverse global degree of coherence, $1/C$, which is linked to the ratio of the single-photon pulse duration to the coherence time \cite{horoshko25}, as a function of $K$. As the pump bandwidth narrows (which in the case of our SPDC source leads to a larger $K$), the relative duration of the central peak narrows relative to the adjacent uncorrelated peaks (cf.\ Fig.\ \ref{fig:g2graphs}a vs \ref{fig:g2graphs}b), showing excellent agreement with the theoretical model \cite{horoshko25}.

\section{Outlook and applications}

We present a new approach enabling observation of maximum bunching for spectrally broadband multimode thermal light, thanks to improving the temporal resolution of $g^{(2)}(\tau)$ measurement by a factor of $660$ using temporal magnification based on an electro-optic time lens. We experimentally demonstrate the theoretical prediction of maximum bunching for multimode thermal signals at sufficiently short timescales.

By demonstrating temporal imaging of signals from spontaneous parametric down-conversion, a process that is used in quantum optical experiments to generate complex quantum states of light, including entangled photon pairs \cite{tsang06} or squeezed light \cite{Patera18}, our results pave the way towards experimental quantum temporal imaging, i.e., temporal imaging of non-classical states of light \cite{Bromberg10},  study of multiphoton correlations or towards quantum sensing applications \cite{Mukamel21}. Furthermore, in the strict quantum optics context, we highlight, in a concrete experimental setting, the important distinction \cite{horoshko25} between the time-resolved autocorrelation at zero delay, $g^{(2)}(\tau = 0)$ and the integrated (pulsed) $g^{(2)}_\mathrm{int}$ \cite{christ11}.   
Beyond quantum state characterization, pre-detection temporal imaging establishes a universal platform for statistical optics across diverse physical systems where ultrashort correlation times previously prevented direct detection. While temporal photon statistics are pivotal in platform-specific regimes—such as subcycle THz field correlations \cite{BeneaChelmus16}, microwave thermal fields in superconducting circuits \cite{Goetz17}, or superfluorescence in perovskite superlattices \cite{Raino18} -- our approach provides a field-agnostic route to resolve $g^{(2)}(\tau)$ directly in the time domain. Furthermore, applying temporal magnification prior to detection opens new avenues for stellar intensity interferometry on broadband starlight and low-light fluorescence spectroscopy in biophysics.

\section*{Methods}

\subsection{Source of pulsed thermal light}
We pumped a 15-mm-long periodically-poled potassium titanyl phosphate crystal (PPKTP) with pulses at 781~nm central wavelength, obtained by second-harmonic generation from an $80$~MHz repetition rate amplified output of an erbium femtosecond oscillator (Menlo Systems C-Fiber HP), to produce collinear, type-II, spontaneous parametric down-conversion (SPDC). The co-propagating orthogonally polarized signal and idler beams with central wavelengths of 1559.5 nm and 1564.5 nm were coupled into a polarization-maintaining single-mode optical fibre and split using a fibre polarizing beam splitter. The idler was discarded without detection for the second-order correlation measurements, allowing us to observe thermal light in the signal beam path \cite{Ivanova06, christ11, Shi20}. Discarding one photon of the twin-beam state yields a chaotic thermal state governed by Gaussian field statistics with a theoretical zero-delay autocorrelation $g^{(2)}(0) = 2$  \cite{Goodman}. 

We achieved high separation of the signal and idler photons by using the source in the non-degenerate configuration and using a home-built telecom-wavelength femtosecond pulse shaper as an adjustable spectral filter (SF) to discard the idler photons, while preserving the signal photons. Additionally, the spectral filter suppressed the side lobes of the SPDC phase-matching function, enabling operation close to the single-mode regime ($K\approx 1$). The fibre-coupled signal field was directed to the temporal imaging setup. 

Note that unlike classical thermal light, increasing the pump bandwidth in non-degenerate SPDC relaxes energy-conservation constraints, which can be used to reduce the frequency correlations between signal and idler photons and to drive the mode number $K$ toward unity ($K \approx 1$) \cite{Grice2001, URen2004}.

\subsection{Temporal imaging}
The signal was directed to a temporal magnification setup \cite{kolner89, Karpinski21}, comprising $0.4$~km of SMF-28 optical fibre (introducing group delay dispersion, GDD, of $-9.64$~ps$^2$), a waveguided lithium-niobate electro-optic phase modulator (EOSpace) driven by a single-tone $16$~GHz, $33$~dBm radio-frequency signal (generated using a $92$~GS/s arbitrary waveform generator, Keysight, synchronized to a reference clock signal derived from a photodiode monitoring the pulses picked off from the femtosecond oscillator \cite{Sosnicki23}) approximating a parabolic time-dependent voltage \cite{kolner89, Karpinski21, Kapoor2025}, and a chirped-fibre-Bragg-grating dispersion compensating module (Proximion AG) introducing GDD of $6361$~ps$^2$. The ratio of GDDs yields the temporal magnification of $660$ for this temporal imaging setup (See Supplementary Section S2 for details on determination of the magnification factor). The resolution of temporal imaging depends on the RF signal amplitude driving the electro-optic phase modulator and equals $3.4$~ps (which was independently measured by temporal imaging of two approximately Gaussian pulses with decreasing separation and using the Rayleigh criterion).

\subsection{Joint spectral intensity measurement and Schmidt number estimation}
We measured the joint spectral intensity by directing the signal and idler beams to two single-photon-counting dispersive Fourier transform (DFT) spectrometers \cite{goda2013, davis17}, Figure \ref{fig:setupTI}(d). In DFT, a medium with large GDD is used to map the spectrum of a pulsed signal to time. We used $10$ and $9.7$~km-long SMF-28 fibre spools for the signal and idler, respectively, followed by the superconducting single-photon detectors and coincident time-tagging (within a $12.5$~ns coincidence window).

The Schmidt number was estimated from the JSI measurement by the singular value decomposition of the joint spectral amplitude (JSA) resulting from the measured JSI with an assumed flat spectral phase.

Please see Supplementary Section S1A for further details.

\subsection{Autocorrelation function calculation}

In the regime where the detectors resolve single photons, Eq. (\ref{g2}) can be rewritten as \cite{Kimble77}
\begin{equation}\label{eq:g2def}
g^{(2)}(\tau) = \frac{P_{s_1, s_2}(t,t+\tau)}{P_{s_1}(t)\cdot P_{s_2}(t+\tau)},
\end{equation}
where $P_{s_x}(t+\tau)$ is the probability of detecting a photon at the signal no.~$x$ detector, where the signal detection is delayed by $\tau$, and $P_{s_1, s_2}(t,t+\tau)$ is the probability of simultaneous detection of two photons at the signal no. 1 and signal no. 2 detectors, where the second signal detection is delayed by $\tau$. Since the detected light is pulsed, $P_{s_x}(t+\tau)$ is inherently time-dependent and approaches zero between consecutive pulses. Consequently, a direct normalization of the second-order coherence function can lead to division by values close to zero and, therefore, to numerical instabilities. To avoid this issue, we use a constant normalization factor determined from the height of the non-central peaks that do not carry information on the number of modes.

\subsection{Data analysis}

The experimental data consist of time tags (timestamps) corresponding to individual photon arrival events, i.e. detector clicks, and the associated detector number, identifying each photon as signal A or B. The data was collected over a continuous 7-day experiment, generating approximately 1.5~TB of binary data. To compute the second-order correlation functions, we constructed histograms of time differences between time tags from signal A and signal B, using bin sizes of either $100$~ps or $12.5$~ns. For normalization, histograms were computed over a range of $\pm 3$~$\mathrm{\mu}$s, yielding $480$ distinct peaks. Each peak was individually fitted with a Gaussian function parametrized by its width, height, and position. The heights of the non-central peaks were averaged and used to normalize the experimental data, yielding the $g^{(2)}(\tau)$ function, see Supplementary Section S3. The $g^{(2)}(0)$ value was then determined as the ratio of the central peak's height to the average height of the non-central peaks. The uncertainty in $g^{(2)}(0)$ was estimated using the standard error of the central peak height fit and the standard deviation of the non-central peak heights. Similarly, the coherence time $\Delta \tau_M$ was found from the central peak width $\Delta t_{M,\mathrm{c}}$ and the average of off-central peaks $\Delta t_{M}$ by deconvolving two Gaussian functions $\Delta \tau_{M} = \left( \Delta t_{M,\mathrm{c}}^{-2} - \Delta t_{M}^{-2} \right)^{-\frac{1}{2}}$ .

\begin{acknowledgments}
We thank Benjamin Brecht for insightful discussions.
This work was supported by QuantERA network, which has received funding from the European Union's Horizon 2020 research \& innovation programme under grant agreement no.~731473, under project ``Quantum information and communication with high-dimensional encoding'' (QuICHE), funded by Agence Nationale de la Recherche, France, grant ANR-19-QUANT-0001 and by National Science Centre, Poland (NCN, Grant No.~2019/32/Z/ST2/00018) and in parts by NCN Grants No. 2019/35/N/ST2/04434 \& 2023/50/E/ST2/00703 and by Bundesministerium f\"ur Forschung, Technologie und Raumfahrt (Germany) via Grant No. 13N16935.
\end{acknowledgments}
\bigskip

\renewcommand\thesection{S\arabic{section}}
\renewcommand\thefigure{S\arabic{figure}}
\setcounter{section}{0}
\setcounter{figure}{0}
\clearpage

\begin{center}
\textbf{\large Supplemental Materials: Observation of Photon Bunching in Broadband Multimode Thermal Light Enabled by Temporal Imaging}
\end{center}

\section{Dispersive spectrometry of correlated photon pairs}

We characterized spectro-temporal correlations of photon pairs by measuring their joint spectral intensity (JSI) via two-photon dispersive spectrometry with the setup shown in Fig.~\ref{fig:setupTI}~(c,e) in the main text. The photon pairs were generated in a periodically-poled potassium titanyl phosphate (PPKTP) crystal waveguide pumped with pump pulses shaped by a home-built femtosecond pulse shaper PS, enabling control of signal-idler correlation in the type-II spontaneous parametric down-conversion (SPDC) process. The signal and idler photons were separated by a fiber polarization beamsplitter (FPBS), with the signal photon additionally passing through the adjustable spectral filter SF, as in the main experiment.

For JSI measurement both photons were coupled into two separate SMF-28 fiber spools with lengths of 10~km and 9.7~km for the signal and idler, respectively, introducing about 180~ps/nm of group delay dispersion. The dispersion maps their wavelengths to arrival times, which we measured with superconducting nanowire single-photon detectors (SNSPDs, Single Quantum) followed by a time-to-digital converter (Swabian TimeTagger Ultra). Coincidence counts yield a 2D arrival-time histogram of the JSI.

\begin{figure}[!t]
    \centering
    \includegraphics[width=\linewidth]{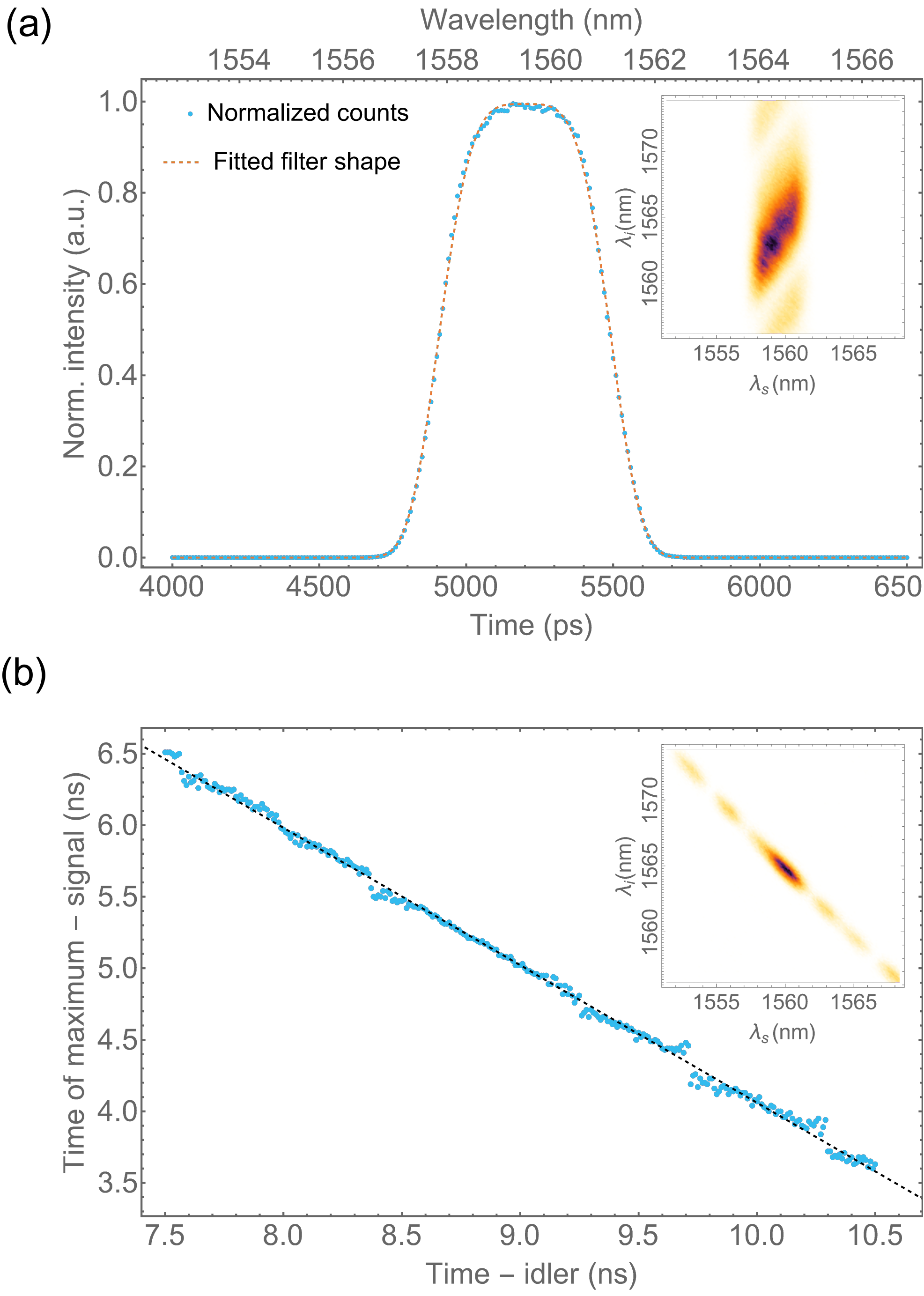}
    \caption{(a) Marginal spectrum of the signal photon (blue points) together with the imposed super-Gaussian spectral profile (dashed orange line). The inset shows the corresponding joint spectral intensity (JSI) measured for this configuration. (b) Time-of-arrival correlation between the signal and idler photons (blue points) for SPDC pumped by a spectrally narrowband pump. The dashed black line shows a linear fit to the measured correlation, from which the idler time-to-wavelength mapping is obtained. The inset shows the corresponding JSI measured in this configuration.}
    \label{fig:s1}
\end{figure}

\begin{figure}[!t]
    \centering
    \includegraphics[width=\linewidth]{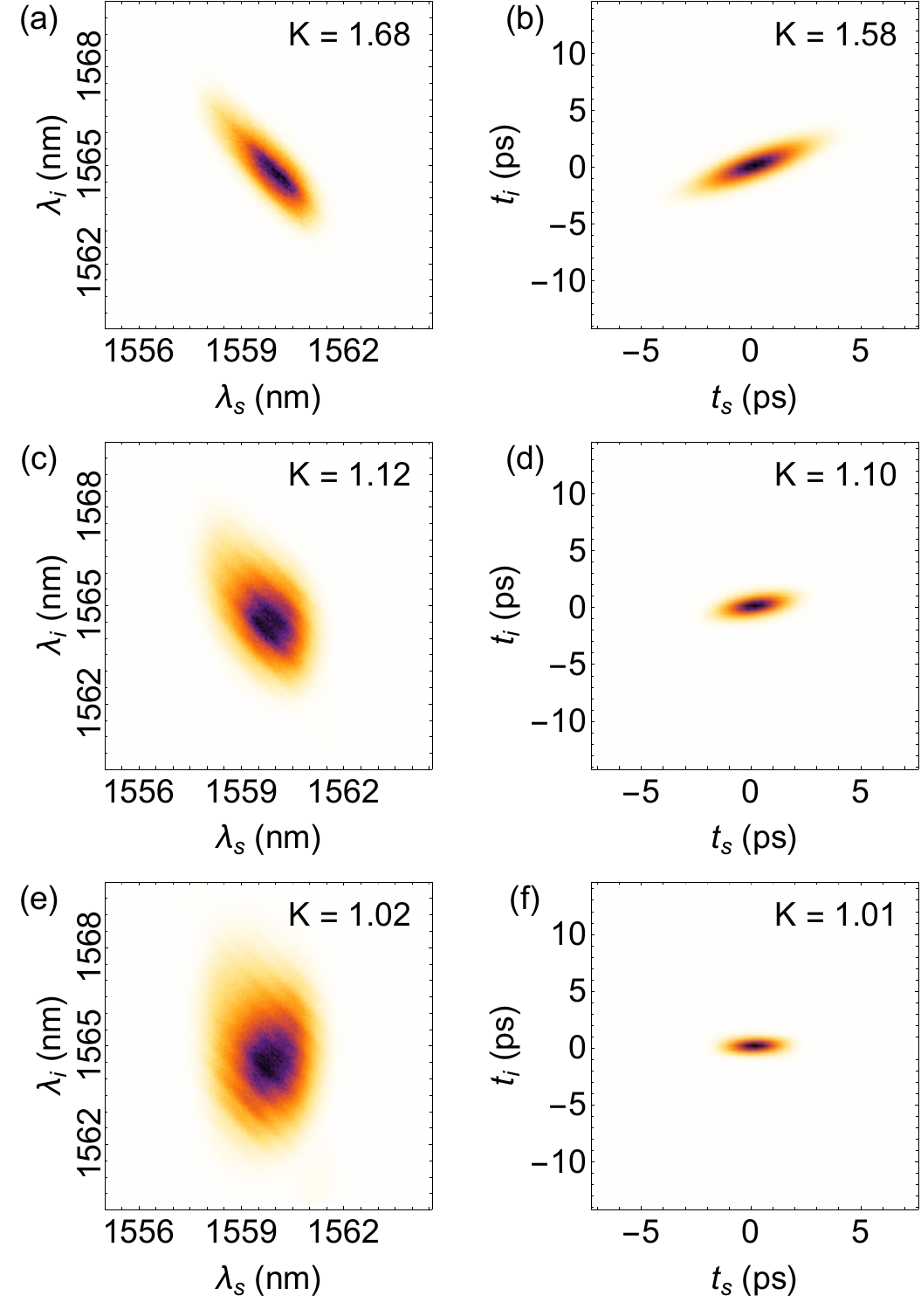}
    \caption{Representative measured joint spectral intensities (JSIs) (a,c,e) and the corresponding joint temporal intensities (JTIs) (b,d,f) for pump spectral bandwidths of 0.20~nm, 0.56~nm, and 0.90~nm, respectively. Varying the pump spectral bandwidth allows the effective number of modes, $K$, to be tuned from 1.02 to 1.68.}
    \label{fig:s2}
\end{figure}

\subsection{Wavelength mapping calibration}

We calibrate the time-of-arrival-to-wavelength mapping to characterize both the spectral-temporal correlations and to verify the magnification of our temporal imaging system. To do this, we first fully open the pump pulse shaper PS and apply a known spectral shape (a flat top super-Gaussian with FWHM of 3~nm) to the signal photons using the adjustable spectral filter SF. We then measure the JSI of the photon pairs, which yields the marginal spectrum of the signal photons that we fit to the imposed spectral profile, as shown in Fig.~\ref{fig:s1}~(a). It provides an absolute calibration between arrival time and wavelength for the signal photons.

Since there is no additional pulse shaper or spectral filter for the idler photons to perform an absolute calibration as for the signal photons, we used energy conservation instead, along with the previously derived calibration for the signal photons. Specifically, we fully open the signal's adjustable spectral filter SF and set a narrow Gaussian filter (0.2~nm FWHM) in the pump pulse shaper PS. This configuration yields a strongly anticorrelated JSI, as dictated by energy conservation, see the inset of Fig.~\ref{fig:s1}~(b). By identifying pairs of arrival times for signal and idler photons and performing a linear fit on these pairs, as shown in Fig.~\ref{fig:s1}~(b), we can relate the wavelengths (and arrival times) of signal and idler photons. By combining it with the previously determined mapping for signal photons, we obtain an absolute calibration between arrival time and wavelength for the idler photons. 

\subsection{Controlling spectral-temporal correlations of photon pairs} 

Using the previously calibrated dispersive spectrometry mapping, we measured JSIs for photon pairs generated with pump bandwidths ranging from 0.1~nm to 1.1~nm in 0.02~nm increments. Representative cases for pump bandwidths of 0.20~nm, 0.56~nm, and 0.90~nm are shown in Fig.~\ref{fig:s2} (a), (c), and (e), respectively. In all experiments we kept the flat-top super-Gaussian profile in the signal's adjustable spectral filter SF to remove the phase-matching side lobes, enabling near-single-mode operation with a pump bandwidth of 0.9~nm. We quantify the effective number of modes $K$ (Schmidt number) via singular value decomposition, assuming a flat spectral phase for each measured JSI. In panels Fig.~\ref{fig:s2} (b), (d), and (f), we also show joint temporal intensity (JTI) calculated based on the 2D-Fourier transform of the joint spectral amplitude (JSA) under the same flat-phase assumption.
 \smallskip

\section{Temporal magnification factor}

The magnification factor of our temporal imaging setup is given by the ratio of the secondary dispersion $D_{2}$ to the primary dispersion $D_{1}$ as $M = - \frac{D_2}{D_1}$. As the primary dispersion medium, we use a 400~m-long SMF-28 fibre with a dispersion of $-9.64$~$\mathrm{ps}^2$ ($7.46$~ps/nm). The secondary dispersion is provided by a dispersion-compensation module (Proximion DCM) in the form of a chirped fibre Bragg grating (CFBG) with a dispersion of 6361~$\mathrm{ps}^2$ (-4927~ps/nm). This yields a temporal magnification factor of $M=660$.

To experimentally verify the magnification factor, we compare the signal photon duration inferred from the JSI measurements, obtained via the temporal widths of calculated JTIs, with temporally magnified signal photons measured directly.

\begin{figure}[!b]
    \centering
    \includegraphics[width=\linewidth]{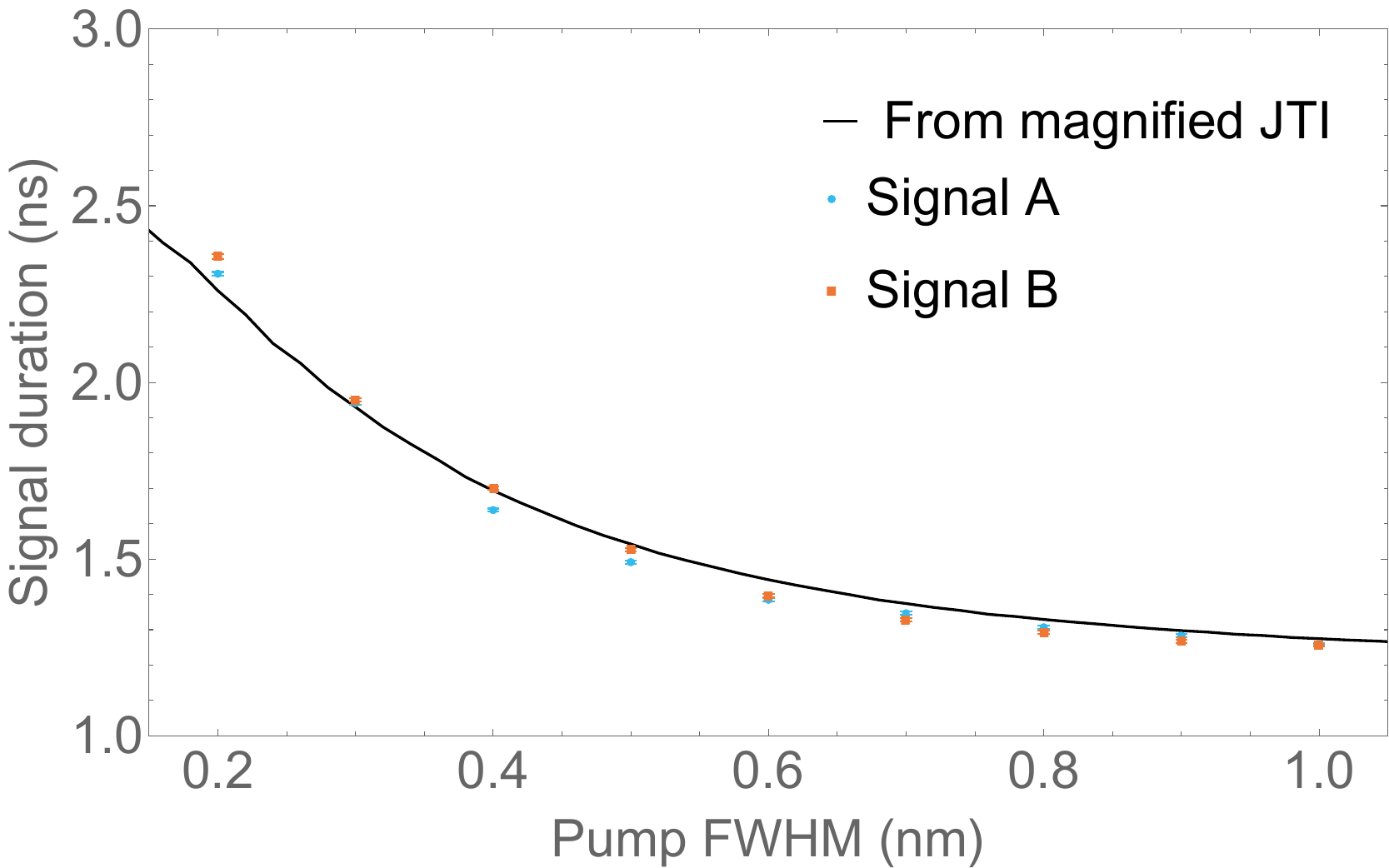}
    \caption{Measured temporally magnified signal-photon durations (FWHM) (blue and orange points) compared with the corresponding durations obtained from the temporal marginal of the JTI and scaled by the temporal magnification factor $M = 660$. }
    \label{fig:s3}
\end{figure}

First, we measured the JSI over pump bandwidths ranging from 0.10~nm to 1.1~nm and used a 2D-Fourier transform and the flat spectral phase assumption to calculate the JTIs for each pump bandwidth, with three exemplary JTIs shown in Fig.~\ref{fig:s2}~(b, d, f). From these, we extract the FWHM duration of the signal marginal and plot the resulting durations, rescaled by $M=660$, as the solid black curve in Fig.~\ref{fig:s3}.

To determine the duration of the temporally imaged signal photons, we record arrival times referenced to the experiment’s clock, defined by the pulsed laser source, during the $g^{(2)}(\tau)$ measurement with the temporal imaging system in place. We obtain two datasets (A and B) by using a 50:50 beamsplitter for Hanbury Brown-Twiss (HBT) interferometry. We construct histograms of the relative arrival times and extract their FWHMs, which we show in Fig.~\ref{fig:s3} as blue and orange points. The resulting signal photon duration shows excellent agreement with the value inferred from the JSI analysis, consistent with the expected magnification factor $M=660$. 

\begin{figure*}[!t]
    \centering
    \includegraphics[width=\linewidth]{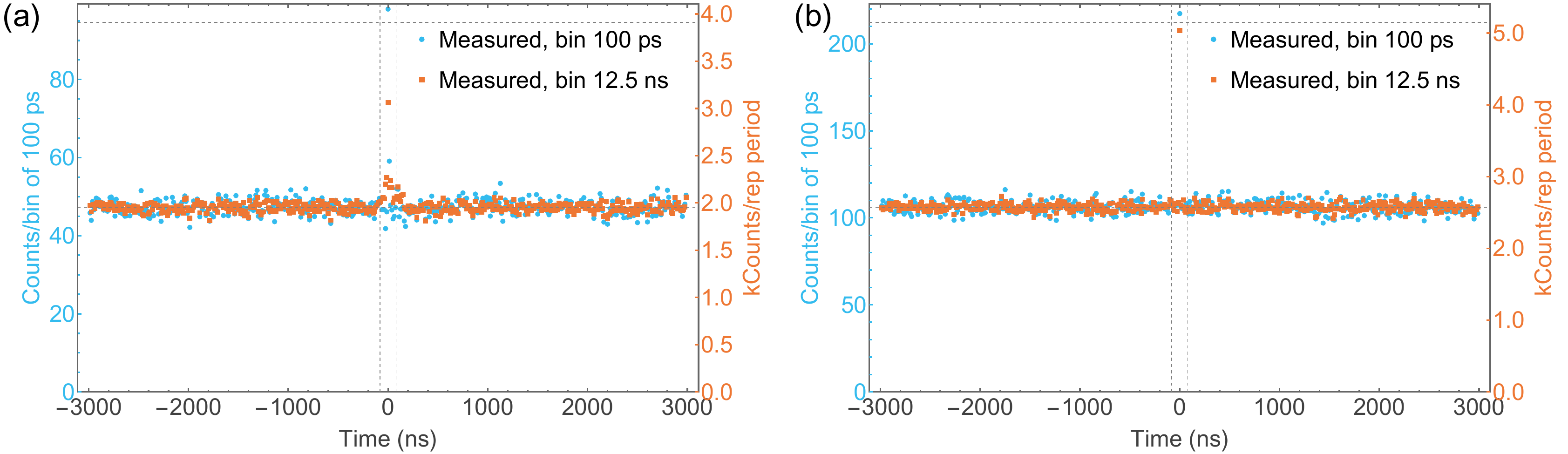}
    \caption{
        Full measurement range of $\pm 3\;\mathrm{us}$ of time-magnified photon count histograms presented in Fig.~\ref{fig:g2graphs}, when pumping the SPDC with a spectral bandwidth of (a) 0.2~nm and (b) 0.9~nm, corresponding to Schmidt numbers $K=1.68$ and $K=1.02$, respectively. The blue points show the maxima of fitted Gaussians for 100~ps binning, while the orange points show the same measurement data as in Fig.~\ref{fig:g2graphs}, i.e., 12.5~ns binning. We use an average height of all non-central peaks to determine the normalization coefficient. The horizontal dashed lines show values corresponding to $g^{(2)}(\tau) = 1$ (bottom) and $g^{(2)}(\tau) = 2$ (top), while the vertical dashed lines indicate the measurement range of $\pm 80\,\mathrm{ns}$ shown in Fig.~\ref{fig:g2graphs}.
    }
    \label{fig:s4}
\end{figure*}

\section{Normalization of the second-order coherence function}

The normalized second-order coherence function $g^{(2)}(\tau)$ experimentally is given by:
\begin{equation}
g^{(2)}(\tau) = \frac{P_{s_1, s_2}(t,t+\tau)}{P_{s_1}(t)\cdot P_{s_2}(t+\tau)},
\end{equation}
where $P_{s_1,s_2}(t,t+\tau)$ denotes the probability of detecting two photons simultaneously in detectors 1 and 2, with a time delay $\tau$, $P_{s_x}(t)$ denotes the probability of detecting a single photon in detector $x$, while their product, $P_{s_1}(t)P_{s_2}(t+\tau)$, serves as the normalization factor. For stationary processes, such as continuous-wave light, the time dependence can be omitted by setting $t=0$. For non-stationary processes, however, such as pulsed light, the probabilities $P_{s_1}(t)$ and $P_{s_2}(t+\tau)$ remain time-dependent and may vanish when no light pulse is present. This leads to numerical instabilities in the normalized second-order coherence function $g^{(2)}(\tau)$. Therefore, in Fig.~\ref{fig:g2graphs}, we present the histogram corresponding to the unnormalized second-order coherence function, $P_{s_1,s_2}(t,t+\tau)$.

To obtain the normalized value of the second-order coherence function at zero delay, $g^{(2)}(0)$, we use the fact that the second-order coherence function approaches unity at sufficiently long times. We therefore determine the normalization coefficient by averaging the heights of the non-central peaks in the time-magnified histograms and the values of the non-central peaks in the histograms with low temporal resolution, evaluated within the $\pm 3\;\mathrm{\mu s}$ range shown in Fig.~\ref{fig:s4}.

\end{document}